\documentclass{acta}
\usepackage{supertabular,lscape,epsfig}
\usepackage{amssymb}
\usepackage{amsmath}
\usepackage{hyperref}
\usepackage{arydshln}
\usepackage{xcolor}

\SetPages{0}{0}

\SetVol{0}{2021}

\begin{document}
\frenchspacing
\sloppy

\begin{Titlepage}
\Title{Over a century of $\delta$~Ceti variability investigation}

\Author{{\'S}reniawska$^1$, E., Kami{\'n}ski$^2$, K., Kami{\'n}ska, M. K., Tokarek, J., Zg{\'o}rz, M.}{Astronomical Observatory Institute,
Faculty of Physics, Adam Mickiewicz University, S\l{}oneczna 36, 60-286 Pozna{\'n}, Poland\\
$^1$e-mail: ewa.sreniawska@gmail.com\\
$^2$e-mail: chrisk@amu.edu.pl}

\Received{Month Day, Year}
\end{Titlepage}

\Abstract{We present results of a 2014-2018 campaign of radial velocity measurements of $\delta$~Ceti. Combining our determination of pulsation period with historical data we conclude that the most likely explanation of observed changes is the presence of a secondary component with a minimum mass of $1.10 \pm 0.05 M_{\odot}$ on an orbit with a period of $169 \pm 6$ years. Consequently, we revised the intrinsic, evolutionary period change rate to not larger than $0.018 \pm 0.004$ s/century, which is significantly lower than previous estimations and is consistent with evolutionary models of Neilson \& Ignace (2015). We did not find any significant multiperiodic frequencies in radial velocity periodograms such as those reported by Aerts et al. (2006) in photometric data from MOST satellite. Using interferometric angular size of $\delta$~Ceti from JMMC Stellar Diameters Catalogue we determined several physical parameters of the star with bolometric flux method. They turned out to be consistent with most previous deteminations, confirming lower mass and slower evolution of $\delta$~Ceti than obtained using different methods by Aerts et al. (2006) and Neilson \& Ignace (2015).}{Stars: individual: $\delta$~Ceti, Stars: fundamental parameters, Stars: oscillations}

\section{Introduction}
$\delta$~Ceti is one of the longest known variable stars of $\beta$ Cephei type.
It is a V=4.08 star with spectral type of B2IV (Anderson, 2012). For a long time the star was considered to be one of a few monoperiodic pulsators in its class. This view was modified by Aerts et al. (2006) who found several non-harmonic frequencies using high quality space photometry, but it was later questioned by Jerzykiewicz (2007) who reanalysed the same data and concluded that none of these frequencies can be considered significant.

The variability of $\delta$~Ceti was first observed spectroscopically about 120 years ago
by Frost and Adams (1903).
Until about 1965 its period was considered constant
within the measurement uncertainty (van Hoof, 1968).
Lane (1977) was the first to notice observed pulsation period change,
but without any final interpretation.
Subsequent research confirmed that the period is changing (Ciurla, 1979)
and most researchers favored a model with constant rate
( Lloyd \& Pike 1984 , Jerzykiewicz et al. 1988, Jerzykiewicz 1999 ),
although the exact value varied depending on data range and selection criteria.
The latest analysis of $\delta$~Ceti period change was published by Jerzykiewicz (2007)
who concluded that a constant period acceleration can no longer be considered a valid explanation
of all pulsation timing derived from photometric observations between 1952 to 2003.


Several attempts have been undertaken to derive the physical parameters of $\delta$~Ceti.
Aerts et al. (2006) determined stellar parameters employing seismic modeling
of new pulsation modes they identified.
Zorec et al. (2009) used the indirect bolometric flux method and focused on calculating the effective temperature.
Levenhagen et al. (2013) carried out spectral synthesis analysis of selected spectral lines in optical high-resolution spectra.
Neilson and Ignace (2015) used stellar evolution models and computed the rate of change of the period to constrain stellar parameters.
Overall, the derived temperatures are consistent with each other, while the mass, radius and luminosities are not consistent between all estimations.


\section{Observations}

\begin{figure}[!htb]
\begin{center}
\includegraphics[width=0.85\textwidth]{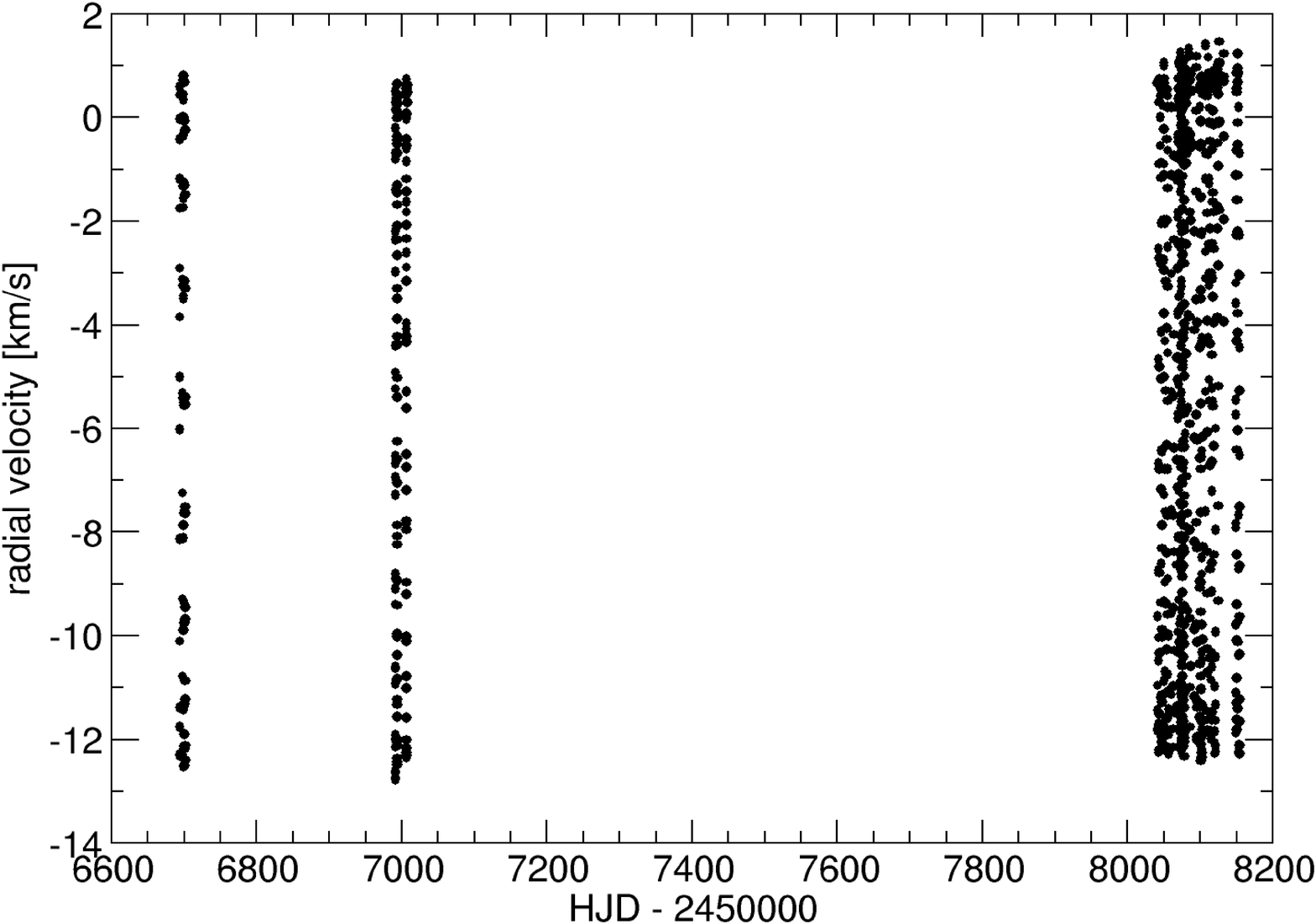}
\FigCap{All 867 heliocentric radial velocities of $\delta$~Ceti measured with RBT/PST2 telescope between 2014 and 2018.}
\end{center}
\end{figure}

\begin{figure}[!htb]
\begin{center}
\includegraphics[width=0.85\textwidth]{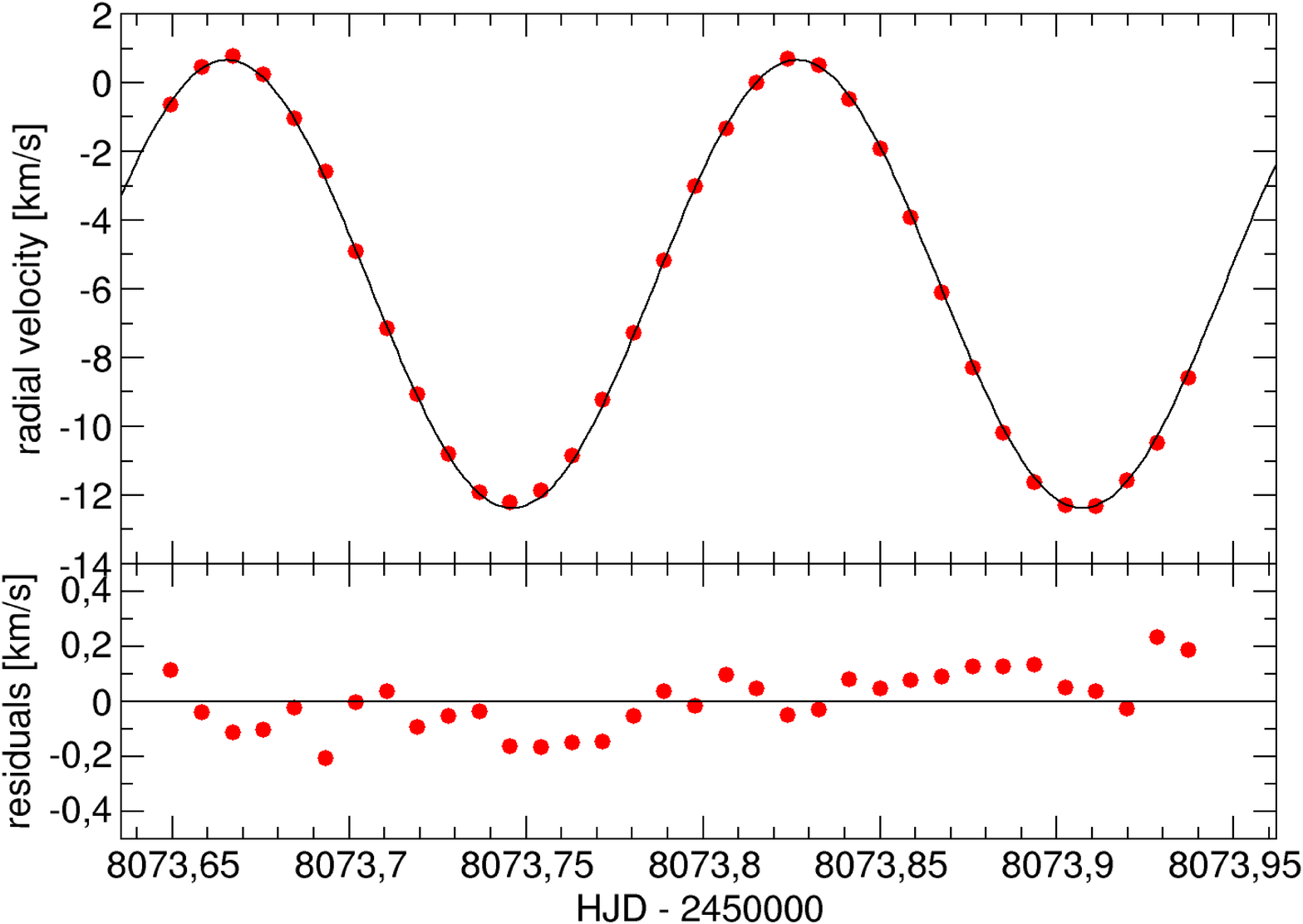}
\FigCap{Radial velocities of $\delta$~Ceti observed on 2017-11-16 with RBT/PST2 telescope. The data are fitted with sine (line) with period of 0.16113762d. Residuals show mostly measurement noise, because the first harmonic is at the level of $SNR<1$.}
\end{center}
\end{figure}


In the years 2014 and 2017-2018 a spectroscopic campaign was carried out
using one of the Global Astrophysical Telescope System telescopes:
Roman Baranowski Telescope/Pozna{\'n} Spectroscopic Telescope 2 (RBT/PST2),
which is located in Winer Observatory in Arizona (Poli{\'n}ska et al., 2014).
RBT/PST2 is a $0.7m$ robotic telescope equipped with $R\approx40000$ fiber-fed echelle optical spectrograph.

Overall, we collected 867 high-resolution spectra in the wavelength range from 3800 to 7100 \AA,
which constitute the largest homogeneous spectroscopic data set for this star (see Fig. 1).

All spectra were reduced using standard IRAF (Tody, 1993) procedures, utilising Th-Ar spectra for wavelength reference and tungsten spectra for flat reduction.

The spectroscopic analysis was performed using the cross-correlation method with IRAF's fxcor tool. Spectrum with the highest signal-to-noise ratio of ${\sim150}$ (recorded on 2014-11-30) was used as a reference. We used the following wavelength ranges for correlation: 4126-4315, 4369-4832, 4904-5190 {\AA} in order to omit telluric and wide stellar lines. We estimate the average radial velocity single measurement uncertainty is at the level of $\sim210$ m/s (see Fig. 2).

\section{Pulsation timing}

Historical data on $\delta$~Ceti pulsation timing has inhomogeneous quality
and some authors restrict their analysis to modern measurements starting from the 1950s onward.
Nevertheless, we decided to use as wide time span as possible
and to combine photometric and spectroscopic observations together.
Photometric measurements were taken from Jerzykiewicz (2007).
Spectroscopic measurements were taken from Lloyd \& Pike (1984),
Jerzykiewicz (1988), Pike \& Lloyd (1991), Kubiak \& Seggewiss (1990) and Aerts et al. (1992).
We rejected spectroscopic measurements of Marshall (1934) and Campos \& Smith (1980)
due to their large residuals with respect to others and
due to only 6 and 8 radial velocity measurements, respectively.

Our radial velocity data was divided into three parts:
from 2014-02-06 to 2014-02-13 (75 measurements),
from 2014-11-30 to 2014-12-16 (134 measurements)
and from 2017-10-14 to 2018-02-05 (658 measurements).
For each part we searched for a phase shift of recorded rv curves maxima
using Period04 (Lenz \& Breger, 2005)
with respect to linear ephemeris from Jerzykiewicz (2007):
\begin{equation}
    \mathrm{HJD}_{\mathrm{lc\; max}} = 2438385.6861 + 0 \overset{\mathrm{d}}{.}16113735 \times E
\end{equation}
where E is the number of cycles from initial epoch.
Our results are the most accurate $O-C$ determinations for $\delta$~Ceti to date (see Tab. 1).

\MakeTable{llllllll}{12.5cm}{Mean epochs of maximum brightness and radial velocity of $\delta$~Ceti.}
{\hline
$HJD - 2400000$ & sigma & type & reference               & $E$       & $O-C$ [d]  \\
\hline
15689.8767  & 0.0056  & rv & Frost and Adams (1903)      & -140848 & 0.0641 \\
23379.9690  & 0.0026  & rv & Henroteau (1922)            & -93124  & 0.0375 \\
26287.8290  & 0.0031  & rv & Crump (1934)                & -75078  & 0.0129 \\
26677.6127  & 0.0059  & rv & Crump (1934)                & -72659  & 0.0053 \\
27096.5685  & 0.0021  & rv & Crump (1934)                & -70059  & 0.0040 \\
34288.9136  & 0.0013  & rv & McNamara (1955)             & -25424  & -0.0165 \\
34630.8475  & 0.0016  & rv & Jerzykiewicz et. al. (1988) & -23302  & -0.0161 \\
34724.6267  & 0.0017  & rv & J{\o}rgensen (1966)         & -22720  & -0.0188 \\
39778.6905  & 0.0006  & rv & Ciurla (1979)               & 8645    & -0.0280 \\
42662.7256  & 0.0012  & rv & Lane (1977)                 & 26543   & -0.0292 \\
43736.5485  & 0.0008  & rv & Lloyd and Pike (1984)       & 33207   & -0.0256 \\
46639.7815  & 0.0002  & rv & Pike and Lloyd (1991)       & 51224   & -0.0042 \\
46744.5135  & 0.0023  & rv & Kubiak and Seggewiss (1990) & 51874   & -0.0115 \\
47109.0177  & 0.0035  & rv & Aerts et al. (1992)         & 54136   & 0.0000 \\
56698.19022 & 0.00010 & rv & this work                   & 113645  & 0.04998 \\
56999.67978 & 0.00010 & rv & this work                   & 115516  & 0.05155 \\
58097.03062 & 0.00005 & rv & this work                   & 122326  & 0.05704 \\
\hdashline
34286.8545  & 0.0022  & lc & Walker (1953)               & -25437  & 0.0192 \\
38338.4765  & 0.0020  & lc & Van Hoof (1968)             & -293    & 0.0036 \\
38385.6861  & 0.0009  & lc & Jerzykiewicz (1971)         & 0       & 0.0000 \\
38656.5605  & 0.0021  & lc & Van Hoof (1968)             & 1681    & 0.0025 \\
39013.9636  & 0.0006  & lc & Jerzykiewicz (1971)         & 3899    & 0.0030 \\
40558.9497  & 0.0020  & lc & Watson (1971)               & 13487   & 0.0042 \\
43061.8957  & 0.0023  & lc & Lane (1977)                 & 29020   & 0.0037 \\
43794.2670  & 0.0021  & lc & Mohan (1979)                & 33565   & 0.0057 \\
43804.7462  & 0.0010  & lc & Jerzykiewicz et. al. (1988) & 33630   & 0.0110 \\
44189.2190  & 0.0010  & lc & Mohan (1979)                & 36016   & 0.0101 \\
44516.9732  & 0.0013  & lc & Sareyan et al. (1986)       & 38050   & 0.0109 \\
44888.7186  & 0.0005  & lc & Jerzykiewicz et. al. (1988) & 40357   & 0.0125 \\
45263.6876  & 0.0010  & lc & Jerzykiewicz et. al. (1988) & 42684   & 0.0149 \\
46744.5534  & 0.0007  & lc & Kubiak and Seggewiss (1990) & 51874   & 0.0284 \\
47008.8208  & 0.0007  & lc & Heynderickx (1991)          & 53514   & 0.0306 \\
48239.2744  & 0.0007  & lc & Jerzykiewicz (2007)         & 61150   & 0.0393 \\
48447.1415  & 0.0016  & lc & ESA (1997)                  & 62440   & 0.0393 \\
52931.7754  & 0.0010  & lc & Jerzykiewicz (2007)         & 90271   & 0.0596 \\
\hline
}

Using the data from Table 1 we fitted four models of $O-C$ changes (Tab. 2).
Each time we also fitted a phase lag of radial velocity curve maximum with respect to light curve maximum.
This lag is slightly model dependent, therefore we could not present all models on a single graph.

The first model (Fig. 3) assumes that $O-C$ changes are parabolic,
which corresponds to the observed pulsation period changing at a constant rate of $0.213$ s/century.
As noted by Jerzykiewicz (2007), it is clear that this interpretation was already 
questionable for data spanning only from 1952 to 2003.
With the data covering years from 1901 to 2018 it is even more evident that this can not
be the only interpretation of all observed changes.

\begin{figure}[!htb]
\begin{center}
\includegraphics[width=0.95\textwidth]{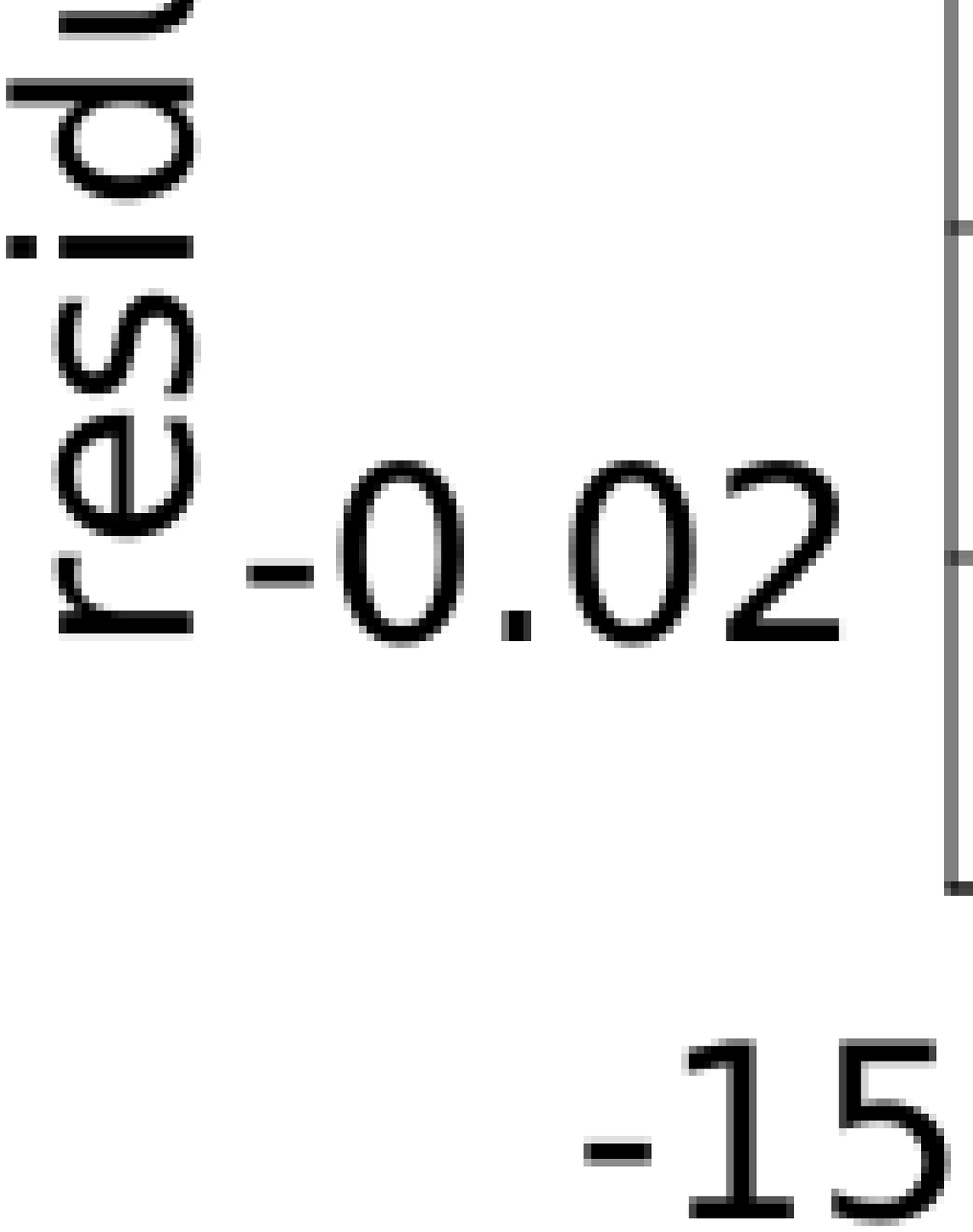}
\FigCap{Observed O-C timing variations of photometric (blue) and spectroscopic (red) maxima with respect to a linear ephemeris with a constant period. Three square points to the right are based on our own spectroscopic measurements. The blue line represents the best fit of model 1, with constant period change. Phase lag of spectroscopic maxima in this case is $0.189$, $\chi^2$ is $5.5 \cdot 10^{-5}$.}
\end{center}
\end{figure}

\MakeTable{lllll}{12.5cm}{Parameters of models fitted to the O-C changes of $\delta$~Ceti observed for 116 years.}
{\hline
                          & model 1              & model 2              & model 3               & model 4 \\
\hline
$\dot{P} ~ [\mathrm{s/century}]$   & $0.213 \pm 0.030$    & -                    & $0.018 \pm 0.004$     & $ 0.220 \pm 0.016$ \\
rv phase  lag             & $0.189 \pm 0.028$    & $0.198 \pm 0.015$    & $0.204 \pm 0.016$     & $ 0.200 \pm 0.016$ \\
$a_1 \sin(i) ~ [\mathrm{au}]$      & -                    & $9.2 \pm 0.6$        & $8.49 \pm 0.34$       & $ 1.36 \pm 0.33$ \\
$e$                       & -                    & $0.29 \pm 0.08$      & $0.34 \pm 0.05$       & $ 0.43 \pm 0.07$ \\
$\omega_1 ~ [\deg]$       & -                    & $306 \pm 8$          & $306 \pm 7$           & $ 89 \pm 8$ \\
$P_{orb} ~ [\mathrm{y}]$           & -                    & $169 \pm 6$          & $169 \pm 6$           & $ 34.3 \pm 1.0$ \\
$T_p ~ [\mathrm{HJD}]$             & -                    & $2444073 \pm 796$    & $2444054 \pm 781$     & $ 2460712 \pm 784$ \\
$f(m_2) ~ [\mathrm{M_{\odot}}]$    & -                    & $0.0270 \pm 0.0045$  & $0.0214 \pm 0.0028$   & $ 0.0022 \pm 0.0017$\\
$m_2 \sin(i) [\mathrm{M_{\odot}}]$ & -                    & $1.19 \pm 0.07$      & $1.10 \pm 0.05$       & $ 0.51 \pm 0.12$ \\
$\chi^2$                  & $5.49 \cdot 10^{-5}$ & $2.00 \cdot 10^{-5}$ & $1.92 \cdot 10^{-5} $ & $ 2.98 \cdot 10^{-5} $ \\
\hline
}

The second model explores the possibility that secular changes of observed pulsation period could be caused by long-period orbital motion of $\delta$~Ceti (Fig. 4). It achieves $\chi^2$ smaller by a factor of $2.5$ than the first model and seems to fit the data much better, without any obvious trends left in residuals. Orbital model was first suggested by Lane (1977), but with a conclusion that it is unlikely. Subsequently researchers focused on determination of the rate of period change, with estimations varying from 0.14 (Cuypers, 1986) to 0.47 (Jerzykiewicz, 1999). Ambiguities of these estimates are easy to understand with our orbital model with a period of $169$ years, because different data sets used by different authors covered different phases of orbital motion.

\begin{figure}[!htb]
\begin{center}
\includegraphics[width=0.95\textwidth]{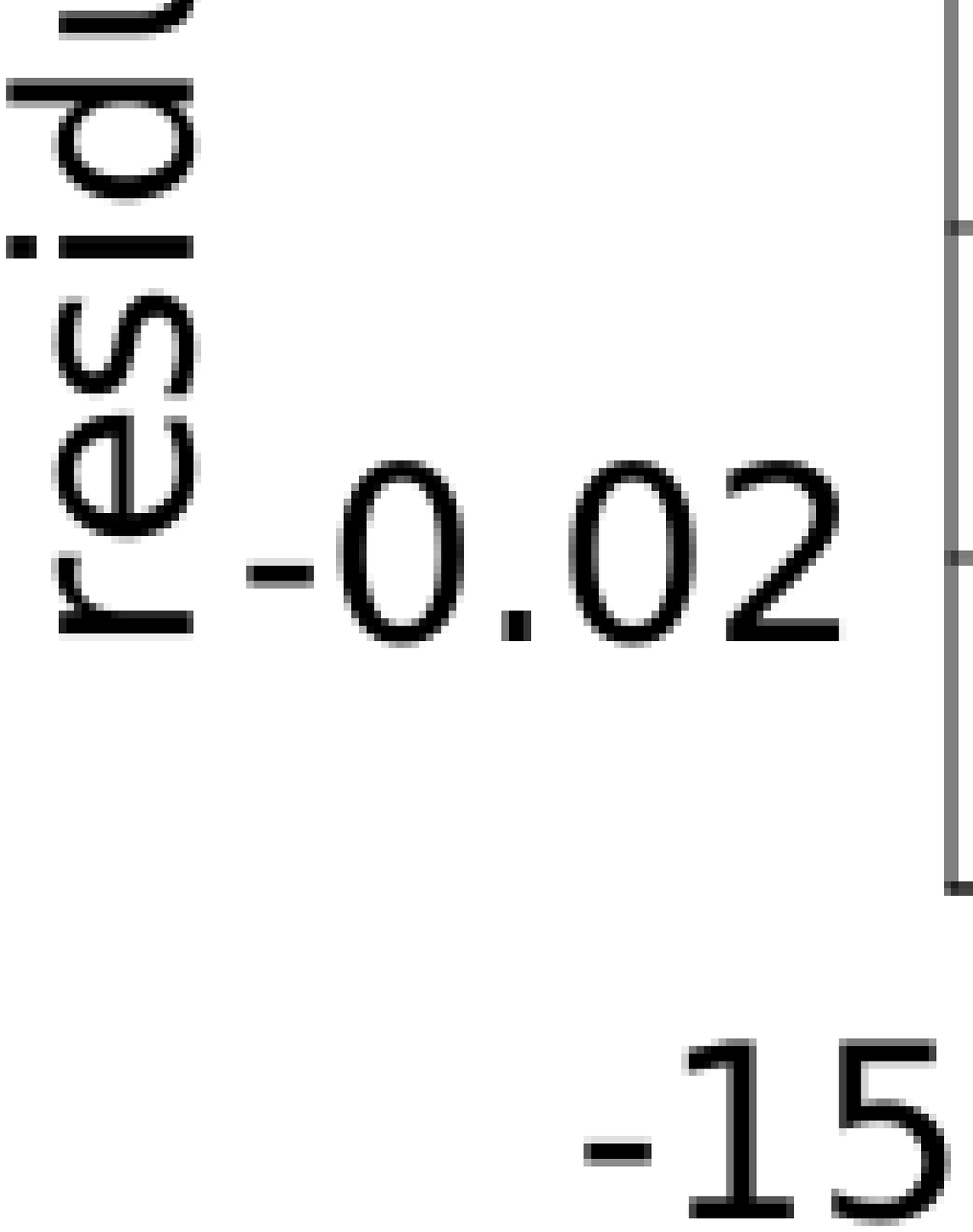}
\FigCap{Observed O-C timing variations of photometric (blue) and spectroscopic (red) maxima with respect to a linear ephemeris with a constant period. Three square points to the right are based on our own spectroscopic measurements. The blue line represents the best fit of model 2 - orbital motion with a period of $169$ years. Phase lag of spectroscopic maxima in this case is $0.198$, $\chi^2$ is $2.0 \cdot 10^{-5}$.}
\end{center}
\end{figure}

The third model adds a parabolic change to the long-period orbital motion from the second model (Fig. 5). This additional free parameter in the model did improve the fit slightly, changing $\chi^2$ per degree of freedom to $1.92 \cdot 10^{-5}$. This makes the third model the most successful of all tested by us. The rate of intrinsic pulsation period changes of $\delta$~Ceti is derived as of $0.018 \pm 0.004$ s/century, about an order of magnitude smaller than suggested previously. Since only about $68\%$ of orbital period is covered by observations, its value should be treated as an early estimate.

\begin{figure}[!htb]
\begin{center}
\includegraphics[width=0.95\textwidth]{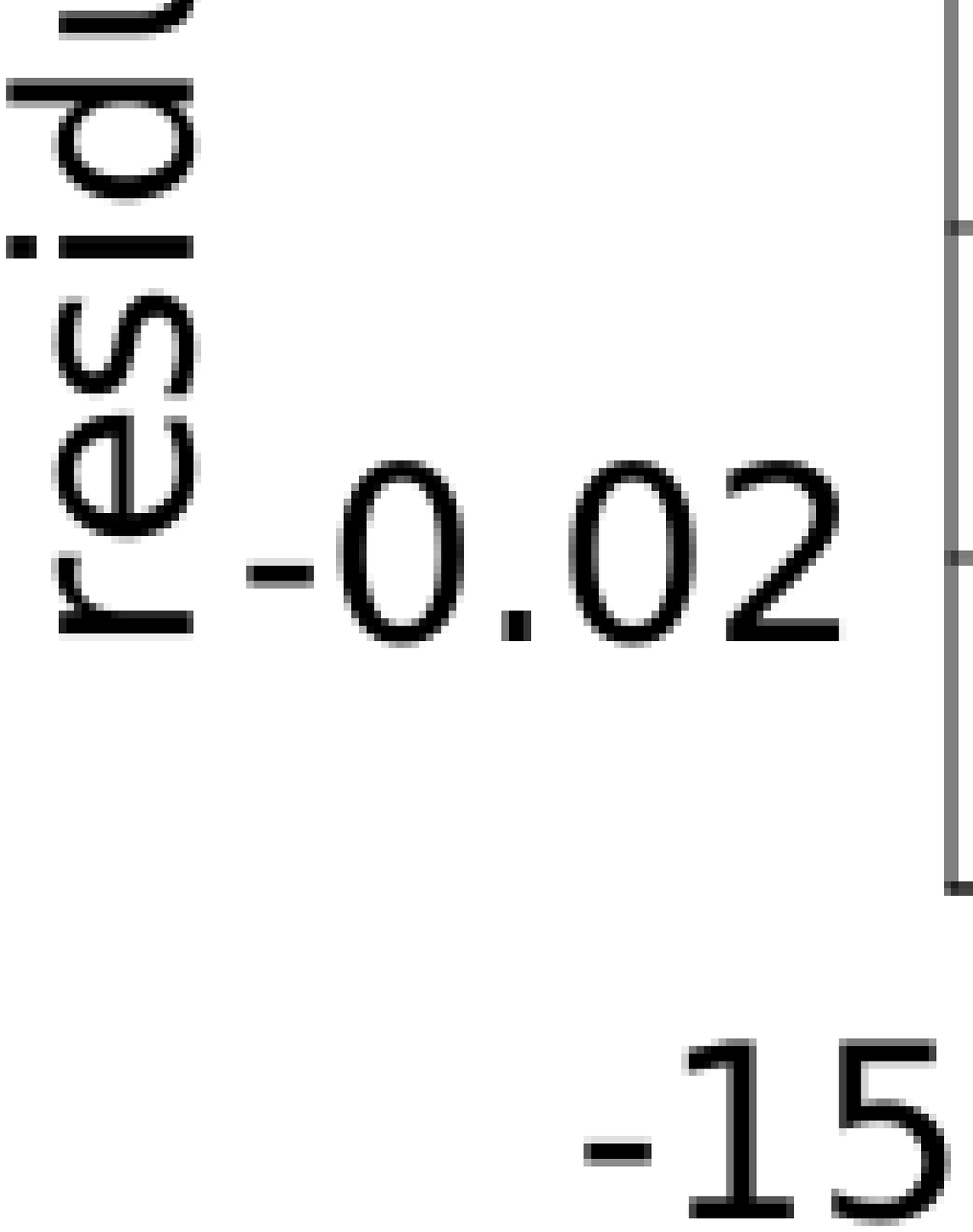}
\FigCap{Observed O-C timing variations of photometric (blue) and spectroscopic (red) maxima with respect to a linear ephemeris with a constant period. Three square points to the right are based on our own spectroscopic measurements. The blue line represents the best fit of model 3 - orbital motion with a period of $169$ years combined with period change at the rate of $0.018$ s/century. Phase lag of spectroscopic maxima in this case is $0.204$, $\chi^2$ is $1.92 \cdot 10^{-5}$.}
\end{center}
\end{figure}

The fourth model assumes that the $O-C$ curve is composed primarily of parabolic change with a short-period orbital motion superimposed (Fig. 6). This concept arises from the systematic variations visible in residuals of pure parabolic model (see Fig. 3). Although a good fit was obtained between $E = -90k$ and $E= + 60k$, there are significant deviations outside this time span, especially with a point at $E \approx 90k$ and our data points at $E \approx 120k$. Moreover, despite the fact that this model uses more degrees of freedom than long-period orbital motion (model 2), it achieves worse $\chi^2 = 2.98 \cdot 10^{-5}$, therefore we consider it an unlikely explanation of the observed $O-C$ variations.

\begin{figure}[!htb]
\begin{center}
\includegraphics[width=0.95\textwidth]{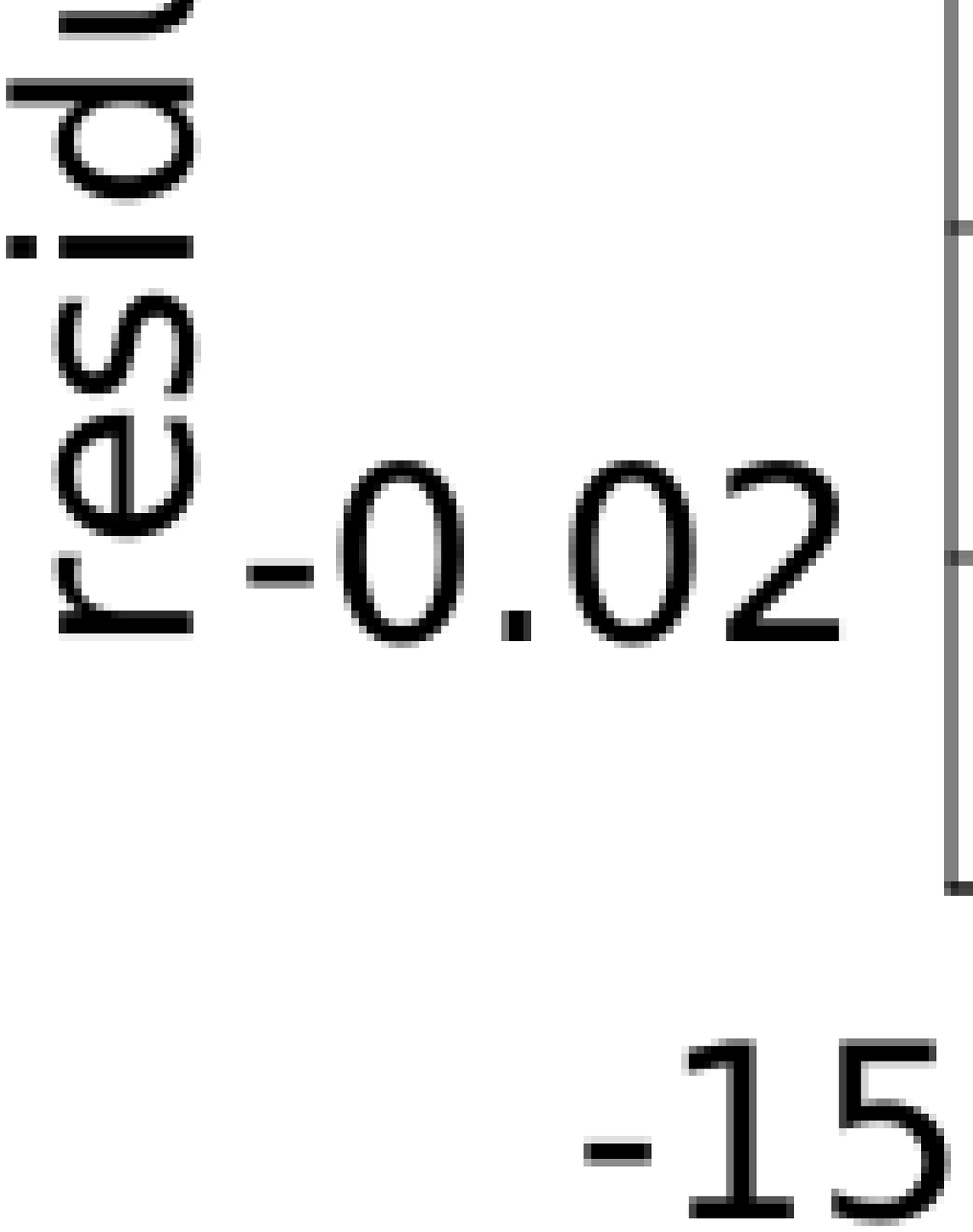}
\FigCap{Observed O-C timing variations of photometric (blue) and spectroscopic (red) maxima with respect to a linear ephemeris with a constant period. Three square points to the right are based on our own spectroscopic measurements. The blue line represents the best fit of model 4 - orbital motion with a period of $34$ years combined with period change at the rate of $0.220$ s/century. Phase lag of spectroscopic maxima in this case is $0.200$, $\chi^2$ is $2.98 \cdot 10^{-5}$.}
\end{center}
\end{figure}

\section{Multiperiodicity}

With our most abundant and dense part of the observing campaign, from \mbox{2017-10-14} to \mbox{2018-02-05}, consisting of 658 rv measurements, we performed a search for multiperiodicity of $\delta$~Ceti. Using Period04 we found only two significant frequencies with ${SNR > 4}$, the primary oscillation period $f_1$ and its harmonic of $f_2 = 2 f_1$ (Tab. 3). The third largest frequency in our sample is $f_3 = 0.052(4)$ $c/d$, but it is at the level of $SNR = 3.5$ which can not be considered as meaningful (Breger et al. 1993). After prewhitening for $f_1$ and $f_2$ the scatter in residuals is at the level of $210$ m/s. Therefore we can not confirm any of the three frequencies found by Aerts et al. (2006) in photometric data from \textit{Microvariability and Oclillations of STars} (MOST) satellite. In the case of the primary $f_1$ frequency our value is slightly smaller than in Aerts et al. (2006) because of secular changes in $\delta$~Ceti observed pulsation period.

\begin{figure}[!htb]
\begin{center}
\includegraphics[width=0.85\textwidth]{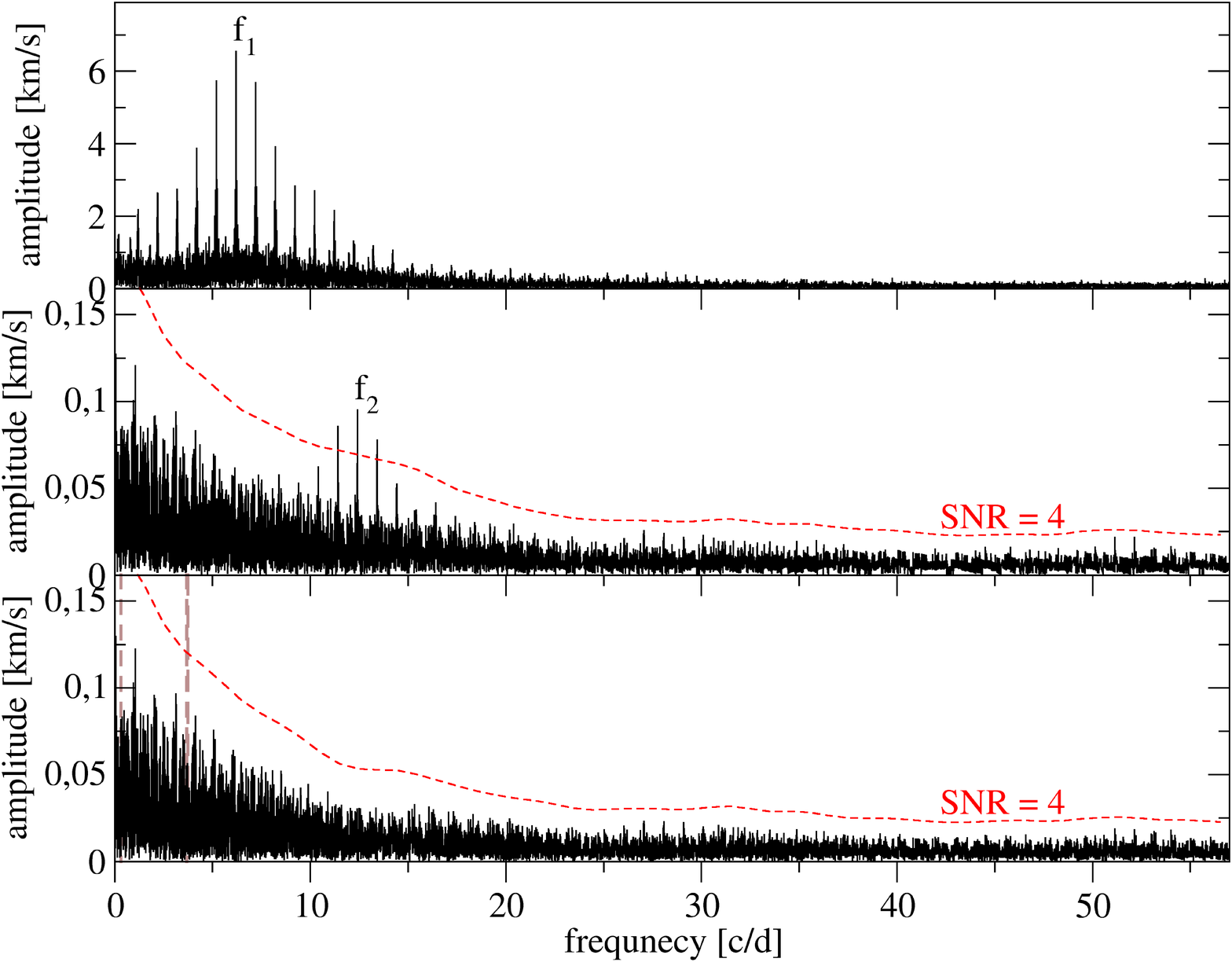}
\FigCap{Periodograms from 658 rv measurements of $\delta$~Ceti. Top figure shows periodogram from raw rv data. In the middle - after prewhitening with $f_1$ frequency. In the bottom, after prewhitening with $f_1$ and $f_2$. Three frequencies found by Aerts et al. (2006) of $3.737$, $3.673$ and $0.318$ $c/d$, are marked in the bottom image with gray vertical dashed lines. No spectroscopic counterparts of Aerts's frequencies are visible.}
\end{center}
\end{figure}

\MakeTable{lllll}{9.0cm}{Radial velocity curve solution.}
{\hline
designation   & frequency [c/d] & amplitude [km/s] & phase shift & $SNR$ \\
\hline
$f_1$         & 6.20586(2)      & 6.5(4)           & 0.357(6) & 276 \\
$f_2 = 2f_1$  & 12.411(6)       & 0.1(0)           & 0.7(4) & 7.0 \\
\hdashline
$f_3$         & 0.052(4)        & 0.1(4)           & 0.5(6) & 3.5 \\
\hline
\multicolumn{5}{p{9.0cm}}{The last frequency is the largest rejected as not significant.
The reference epoch for phase shift is $HJD = 2438385.6861$.}
}

\section{Physical parameters}

Multiple derivations of physical parameters of $\delta$~Ceti in the past relied on
spectrum fitting with LTE or non-LTE models or other indirect methods.
With the new interferometric measurements of angular stellar diameters
and accurate paralaxes it became possible
to derive several physical parameters of $\delta$~Ceti with a minimum number
of assumptions using the bolometric flux method. For the description of the method see
for example Zorec et al. (2009) or Jerzykiewicz and Molenda-{\.Z}akowicz (2000).

We took the apparent magnitudes of $\delta$~Ceti of ${V = 4.08}$ and ${B = 3.87}$
from XHip catalog (Anderson \& Francis, 2012).
The correction for interstellar extinction was based on
coefficients of ${A_V = 0.089}$ and ${A_B = 0.121}$, $E(B-V) = 0.032$,
taken from NED Extinction Calculator\footnote{\url{https://ned.ipac.caltech.edu/extinction_calculator}}
which uses data from Schlafly \& Finkbeiner (2011).
The limb darkened angular diameter of $0.230 \pm 0.021 ~ mas$ was taken
from JMMC Stellar Diameters Catalogue\footnote{\url{http://www.jmmc.fr/jsdc}}
(Bourg{\'e}s et al., 2014).
Using formulas for bolometric corrections from Torres (2010),
we derived for $\delta$~Ceti ${BC_V = -2.09 \pm 0.25}$, bolometric radiation flux
received at the Earth ${f_{bol} = (4.19 \pm 0.93) 10^{-9} ~ Wm^{-2}}$ and finally
the effective surface temperature of $22090 \pm 1580 ~ K$.

Using XHip parallax of ${5.02 \pm 0.15 ~ mas}$
we also derived the star's radius of ${4.92 \pm 0.47 ~ R_{\odot}}$
and luminosity of ${\log L/L_{\odot} = 3.71 \pm 0.10}$.
We did not use GAIA EDR3 catalog because the uncertainty
of its parallax is significantly higher than in XHip catalog.

Additionally, taking $\delta$~Ceti mass of ${7.9 ~ M_{\odot}}$ from Levenhagen et al. (2013)
we calculated the $\log g$ value of ${3.95 \pm 0.09}$.

With the exception of bolometric corrections from Torres (2010)
the method used here is independent of stellar or spectroscopic modeling.
A similar method, although with indirectly inferred stellar angular diameter,
was previously used for $\delta$~Ceti by Zorec et. al (2009). Therefore the presented derivation
is the first independent verification of its physical parameters.
The uncertainty of $T_{eff}$ and $L$ determined with this method
is dominated by the uncertainty of interstellar extinction
and to a lesser degree by the uncertainty of stellar disc angular diameter,
while $\log g$ is dominated strongly by the star's angular size measurements.

Table 4 presents the comparison of our results with previously published
physical parameters of $\delta$~Ceti.
We can see that our temperature is consistent with all other derivations.
In case of $\log g$ and luminosity there are, however, some discrepancies.
Some authors suggested that $\delta$~Ceti is brighter
and with smaller $\log g$ than found by us. This is discussed further in the next section.

\MakeTable{llllllllll}{12.5cm}{Comparison of physical parameters of $\delta$~Ceti}
{\hline
                     & This work         & NI15          & L13     & L10     & H09     & Z09     & A06     & ND05    & GL92 \\
\hline
$T_{eff}$ (K)        & $22090 \pm 1580$  & $20900$-      & $21650$ & $23000$ & $21900$ & $22610$ & $22233$ & $21827$ & $23750$ \\
                     &                   &$26300$\\
$\log g$             & $3.95 \pm 0.09$   & $-$           & $4.03$  & $3.9$   & $4.05$  & $-$     & $3.73$  & $3.73$  & $4.08$ \\

$\log L$/$L_{\odot}$ & $3.71 \pm 0.10$   & $4.2$ - $5$ & $3.6$   & $-$     & $3.6$   & $-$     & $4.06$  & $-$     & $-$ \\
\hline
\multicolumn{10}{p{12cm}}{NI15 - Neilson and Ignace (2015), L13 - Levenhagen et al. (2013), L10 - Lefever et al. (2010), H09 - Hubrig et al. (2009), Z09 - Zorec et al. (2009), A06 - Aerts et al. (2006), ND05 - Niemczura and Daszy{\'n}ska-Daszkiewicz (2005), GL92 - Gies and Lambert (1992).}
}

\section{Conclusions}

We supplemented the 1901-2003 measurements of $\delta$~Ceti pulsation timing available in literature with three new measurements in 2014 and 2017/18 using our 867 spectra recorded with RBT/PST2 telescope. After comparing different models of $O-C$ variation we conclude that the most likely explanation of these changes is that $\delta$~Ceti is a binary system with a period of $169 \pm 6$ years. The secondary component's minimum mass is $1.1 \pm 0.05 ~ M_{\odot}$ and assuming the primary mass of $7.9 M_{\odot}$ (Levenhagen et al., 2013) its relative orbit size is $a_{12} sin(i) = 69 \pm 8 ~ AU$. If the secondary is a main sequence star, then from mass-luminosity relation
we can estimate its minimum luminosity $L_{2} \gtrsim 0.0004 ~ L_{\delta Ceti}$. This implies that the maximum $\Delta V = V_2 - V_{\delta Ceti} \approx 8.5$, which is large enough for the star to be undetectable in spectroscopy. Assuming that the secondary component could be detected at luminosity of $0.01 L_{\delta Ceti}$, its mass can not be larger than about $2.5 M_{\odot}$, therefore the inclination can not be smaller than about $26^{\circ}$.

The angular separation between primary and possible secondary component of $\delta$~Ceti system is quite substantial. For the most probable mass range of $1.1 - 2.5 M_{\odot}$ we get $a_{12}$ from $69$ to $157 AU$ which, taking into account the orbit shape and orientation, corresponds to a maximum possible angular separation from $0.40$ to $1.0$ arcsec. Currently the angular distance between components should be between $0.12$ and $0.8$ arcsec depending on inclination, which makes it possible for direct detection using lucky imaging or adaptive optics. For most inclinations the distance will be decreasing in the following years (Fig. 8).

\begin{figure}[!tbh]
\begin{center}
\includegraphics[width=0.7\textwidth]{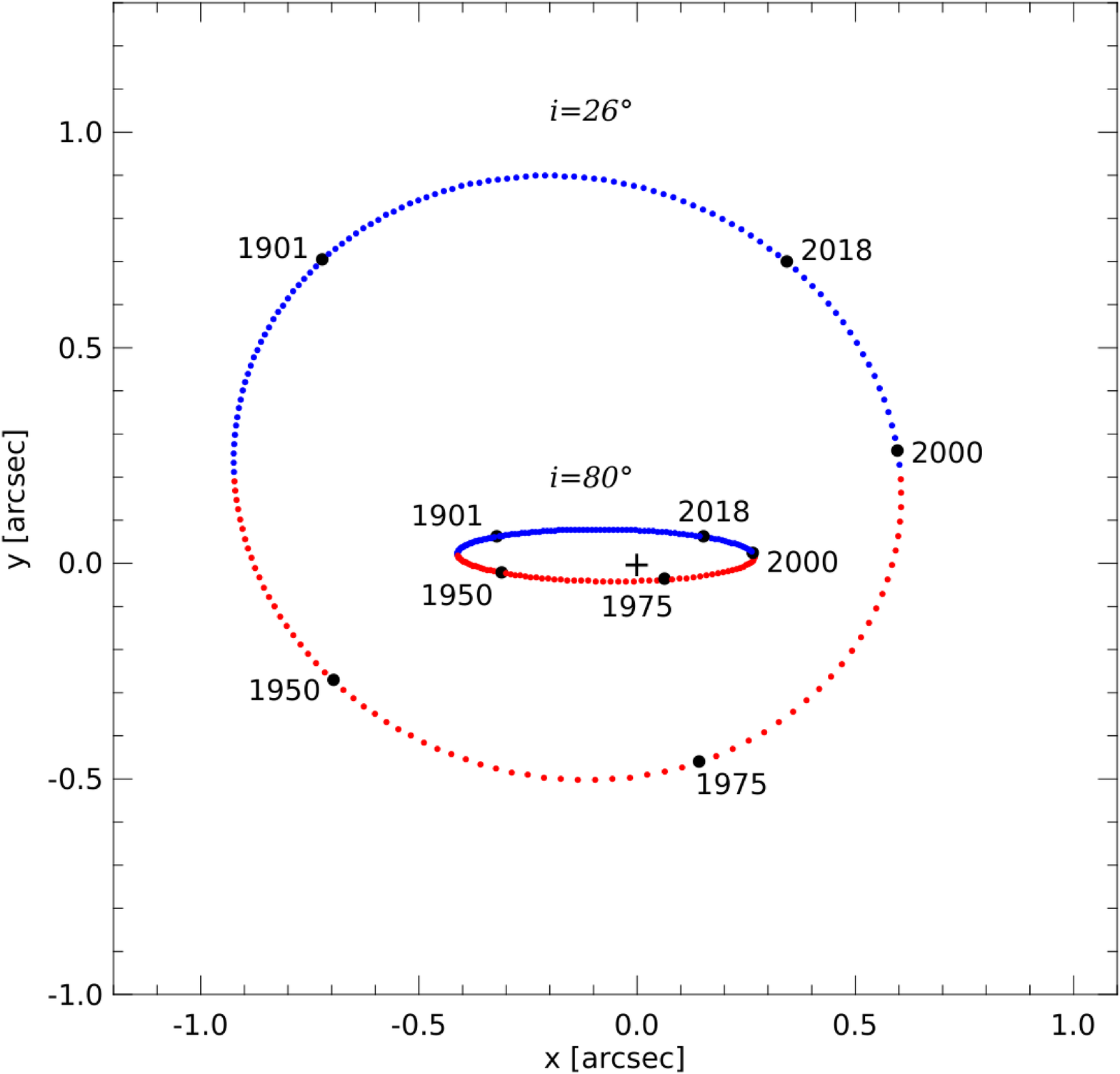}
\FigCap{Position of the hypothetical secondary component (dots) relative to the primary component of $\delta$~Ceti (cross) for two different assumed inclinations. Red dots mark the more distant half of the orbit, blue dots are positions closer to the Earth. Several dates, including the first (1901) and the last (2018) measurement analysed in this paper, are marked. The longitude of ascending node is unknown, so we do not know the position angle of the orbit, therefore arbitrary x and y axes are used.}
\end{center}
\end{figure}

Because of the presence of a secondary component the primary star of $\delta$~Ceti
should be changing its observed proper motion. Comparing astrometric data from XHip catalog 
(Anderson et al. 2012) for epoch of 1991.25
with GAIA EDR3 catalog (Gaia Collaboration et al. 2016, Gaia Collaboration et al. 2021) for epoch of 2016.0
we could not find any significant changes in proper motion (Tab. 5).
According to our estimates during these $24.75$ years the proper motion of $\delta$~Ceti should have changed by about $3.4$ mas/year for $i=90^{\circ}$ or $8.2$ mas/year for $i=26^{\circ}$. This is larger than formal uncertainties of values in both catalogs and therefore the hypothesis of a secondary component remains inconsistent with astrometric data about $\delta$~Ceti. Kervella et al. (2019) using a different method found that there is an inconsistency between Hip2 and GAIA DR2 catalog position of $\delta$~Ceti (at the level of $SNR=3.75$) which might be caused by a companion with a mass of $2.2 \pm 1.0 M_{\odot}$ (assuming $a_2 = 60~AU$).

\MakeTable{llllllllll}{12.5cm}{Comparison of proper motion values for $\delta$~Ceti}
{\hline
source                              & $PM_{RA}$ [mas/y]   & $PM_{Dec}$ [mas/y] \\
\hline
XHip (Anderson et al., 2012)        & $12.85 \pm 0.17$  & $-2.94 \pm 0.11$ \\
GAIA EDR3 (Gaia Collaboration et al. 2021)  & $13.07 \pm 0.31$  & $-2.70 \pm 0.23$ \\
\hline
}

Assuming that the majority of $O-C$ variability can be attributed to orbital motion of $\delta$~Ceti, we found that any remaining secular systematic changes of pulsation periods would be much smaller than previously accepted. With our best model 3, which combines orbital motion and linear pulsation period changes, we found that the intrinsic period change is at the level of $0.018 \pm 0.004$ s/century. If this is correct, then it can solve modeling problems found by Neilson \& Ignace (2015) for $\delta$~Ceti. The authors assumed after Jerzykiewicz (1999) that the period change is either $0.28$ or $0.47$ s/century and concluded that a fast evolving star with a mass of $10-12 M_{\odot}$ and $\log L/L_{\odot} \sim 4.2$ is necessary to explain it. This is inconsistent with most other evaluations of $\delta$~Ceti physical properties, including our (see Tab. 4). With intrinsic pulsation period change rate significantly reduced a lower mass and luminosity seems sufficient for evolutionary models to explain it.

Another recent paper arguing that $\delta$~Ceti is more massive and brighter than our evaluation is Aerts et al. (2006). Their result relies primarily on new pulsation frequencies found in MOST satellite photometry. This result, however, was discussed by Jerzykiewicz (2007) who reanalysed the data and found the new frequencies to have $SNR < 4$ or not be present in the data at all. Our analysis of 658 rv measurements covering time span of 115 days revealed only primary frequency and its harmonic with no signs of any other statistically significant frequencies, which indirectly supports the analysis of Jerzykiewicz and questions the hypothesis of multiperiodicity of $\delta$~Ceti.

We also derived physical parameters of $\delta$~Ceti using the bolometric flux method. Our results of ${T_{eff} = 22090 \pm 1580}$, ${\log g = 3.95 \pm 0.09}$ and ${\log L/L_{\odot} = 3.71 \pm 0.10}$ are consistent with Levenhagen et al. (2013). Although the accuracy of this result is limited by the uncertainty of stellar angular diameter and extinction, this is almost purely geometrical method, relying only on bolometric correction evaluation. This strengthens the idea that $\delta$~Ceti is a less luminous and less massive star than derived by Aerts et al. (2006) and Nielsen \& Ignace (2015).

Finally, we derived the phase lag of rv maximum with respect to light curve maximum. Its value is between $0.189$ and $0.204$, depending on the model we used to fit $O-C$ data. This is slightly smaller than the value of $0.23$ found by Kubiak \& Seggewiss (1990) but consistent with Cyupers (1986) and Ciurla (1979) and Jerzykiewicz et al. (1988) which found the phase lag of $0.2$, $0.19$ and $0.20$, respectively.

\Acknow{
This work has made use of data from the European Space Agency (ESA) mission
{\it Gaia} (\url{https://www.cosmos.esa.int/gaia}), processed by the {\it Gaia}
Data Processing and Analysis Consortium (DPAC,
\url{https://www.cosmos.esa.int/web/gaia/dpac/consortium}). Funding for the DPAC
has been provided by national institutions, in particular the institutions
participating in the {\it Gaia} Multilateral Agreement.
}

\end{document}